\documentclass[%
preprint,
longbibliography,
 amsmath,amssymb,
 aps,
 pra,
]{revtex4-2}

\newcommand{\env}[1]{\texttt{#1}}
\usepackage{graphicx}
\usepackage{dcolumn}
\usepackage{bm}
\usepackage{floatrow}
\usepackage{comment}
\usepackage{mathrsfs}
\usepackage{xcolor}
\usepackage{hyperref}
\usepackage{cleveref}
\crefname{section}{§}{§§}
\Crefname{section}{§}{§§}
\hypersetup{
  colorlinks   = true, 
  urlcolor     = blue, 
  linkcolor    = blue, 
  citecolor   = blue 
}

\begin{document}

\preprint{}

\title{Control of growth of local heat release rate fluctuations to suppress thermoacoustic instability}

\author{Manikandan Raghunathan}\email{manikandanraghu05@gmail.com}\affiliation{Department of Aerospace Engineering, Indian Institute of Technology Madras, Chennai 600036, India}\author{Nitin Babu George}\affiliation{Potsdam Institute for Climate Impact Research, Potsdam, Germany \\ Department of Physics, Humboldt University of Berlin, Berlin, Germany}\author{Vishnu R Unni}\affiliation{Department of Mechanical and Aerospace Engineering, Indian Institute of Technology Hyderabad, 502284, India}\author{J\"urgen Kurths}\affiliation{Potsdam Institute for Climate Impact Research, Potsdam, Germany \\ Department of Physics, Humboldt University of Berlin, Berlin, Germany}\author{Elena Surovyatkina}\affiliation{Space Research Institute of Russian Academy of Sciences, Moscow, Russia}\author{R. I. Sujith}\affiliation{Department of Aerospace Engineering, Indian Institute of Technology Madras, Chennai 600036, India}



\date{\today}
\begin{abstract}
This experimental study investigates the dynamical transition from stable operation to thermoacoustic instability in a turbulent bluff-body stabilized dump combustor. We conduct experiments to characterize the dynamical transition utilizing acoustic pressure and local heat release rate fluctuations. We observe the transition to thermoacoustic instability for these experiments as we decrease the equivalence ratio towards a fuel-lean setting. More importantly, we observe significant growth of local heat release rate fluctuations near the bluff-body well before the appearance of large-scale spatial or temporal patterns during the occurrence of thermoacoustic instability. By strategically positioning slots (perforations) on the bluff-body, we ensure the reduction of the growth of local heat release rate fluctuations at the stagnation zone near the bluff-body at operating conditions far away from the onset of thermoacoustic instability. This reduction in the local heat release rate fluctuations ensures that the transition to thermoacoustic instability is avoided. We find that modified configurations of the bluff-body that do not quench these local heat release rate fluctuations at the stagnation zone results in the transition to thermoacoustic instability. We also reveal that an effective suppression strategy based on the growth of local heat release rate fluctuations requires an optimization of the area ratio of the slots for a given bluff-body position. Further, the suppression strategy also depends on the spatial distribution of perforations on the bluff-body. Notably, an inappropriate distribution of the slots which does not quench the local heat release rate fluctuations at the stagnation zone may even result in a dramatic increase in amplitudes of pressure oscillations.
\end{abstract}
\keywords{Thermoacoustic instability; Growth of local heat release rate fluctuations, Passive control}
\maketitle


\section{\label{Introduction} Introduction}

The catastrophic onset of thermoacoustic instability has been a challenging barrier in developing low emission propulsion and power generation systems. Thermoacoustic instability, which manifests as high amplitude periodic pressure and heat release rate oscillations, causes structural failure and overwhelms the thermal protection system \cite{huang2009dynamics, juniper2018sensitivity}. Thus, investigating the onset of the transition to thermoacoustic instability and mitigating its occurrence has gained substantial scientific interest in the past decades. 

Two main control strategies exist for mitigating thermoacoustic instability: active and passive control. Active control relies on external energy sources employing actuators to keep the combustor from deviating significantly from a desired operating state \cite{docquier2002combustion}. Unlike active control methods, passive control strategies seek to alter the fundamental dynamics of the combustion system and break the mutually enhancing interactions between unsteady heat release rate and chamber acoustics \cite{schadow1992combustion, paschereit1998control, labry2010suppression}. In recent decades, studies have shown that by modifying the spatiotemporal dynamics of the flame/flow through injection of secondary air/fuel, mitigation of thermoacoustic instability is possible \cite{schadow1990multistep,lee2000effect,altay2010mitigation,hussain2019investigating}. Many passive control strategies are based on suppressing the visible large-scale patterns such as large-scale coherent structures, standing waves that characterize thermoacoustic instability. However, such large-scale patterns may appear well after the transition to  thermoacoustic instability has occurred.

Traditionally, the onset of thermoacoustic instability in turbulent combustors has been considered as a sudden transition in the system dynamics wherein the system behaviour shifts from stable operation to unstable operation \cite{lieuwen2002experimental}. However, it has been shown recently that in turbulent combustors the transition is not abrupt but instead occurs through a state of intermittency \cite{nair2014intermittency}, wherein bursts of periodic oscillations occur amidst aperiodic oscillations. Furthermore, measures based on pressure fluctuations derived from recurrence plots and multifractality have shown early detection of the impending onset of thermoacoustic instability \cite{nair2014multifractality,juniper2018sensitivity}. Similarly, measures based on spatial patterns of flame fluctuations such as spatial correlation also indicate the impending onset of thermoacoustic instability \cite{george2018pattern}. These temporal and spatial based measures change well before the appearance of the large-scale spatial or large-amplitude temporal patterns, which suggests that the signs of the impending transition to thermoacoustic instability appears much earlier. Therefore, preventing the transition to thermoacoustic instability requires suppressing the onset of the transition itself, which occurs far ahead of large-scale patterns that appear during thermoacoustic instability. 

Recently, in a bluff-body stabilized dump turbulent combustor, Raghunathan et al. \citep{manikandan2021seeds} showed early local growth of flame fluctuations prior to the onset of thermoacoustic instability. In particular, they revealed significant growth of local heat release rate fluctuations occurring at certain zones within the combustor, located around the bluff-body. Additionally, they revealed interconnections that emerged between these zones prior to the transition to thermoacoustic instability. In the present study, we target one of these interconnected zones. We perform experiments with modified designs of the bluff-body to reduce the growth of local heat release rate fluctuations at a critical location revealed in \citep{manikandan2021seeds}, namely the stagnation zone upstream of the bluff-body. Reducing the growth of local heat release rate fluctuations at the stagnation zone results in the suppression of the transition to thermoacoustic instability.

\section{Experimental setup}
\label{2. Experimental Setup}

\begin{figure*}[t]
\centering
\includegraphics[width=0.9\textwidth]{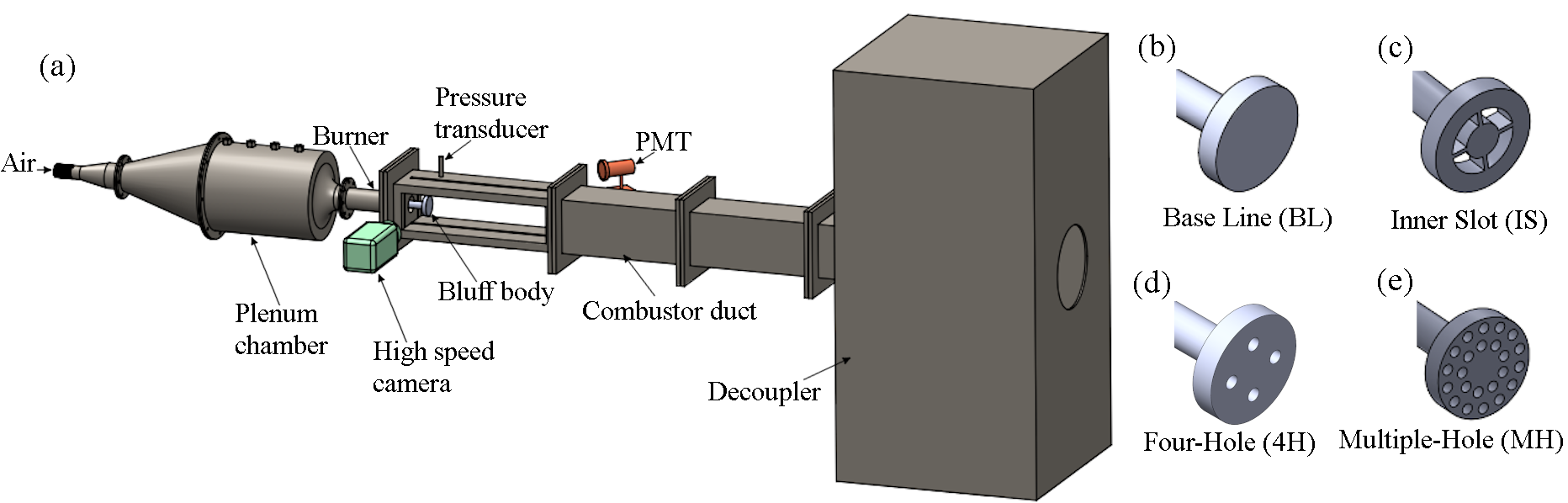}
\caption{Schematic of the experimental setup along with the different bluff-body designs used in this study. a) The main components of the setup are a plenum chamber, a combustion duct and a decoupler. The data acquisition system comprises of a piezoelectric pressure transducer, a photomultiplier tube and a high-speed camera. We use four different designs of the bluff-body in the study, namely, b) Baseline (BL), c) Inner Slot (IS), d) Four-Hole (4H) and e) Multiple-Hole (MH).}
\label{Expt rig}
\end{figure*}

We perform experiments in a bluff-body stabilized dump turbulent combustor, shown in figure \ref{Expt rig}. Air enters through the plenum chamber while fuel enters and mixes with air just before the burner, resulting in partially premixed reactants. In all our experiments, we use Liquefied Petroleum Gas (LPG) as the fuel. The flame is stabilized using a bluff-body, which is a circular disk of 47 mm diameter and 10 mm thickness. The cross-section of the combustion chamber is 90 mm $\times$ 90 mm with a combustor length of 1100 mm. We perform experiments at different bluff-body position from the dump plane: 40 mm and 45 mm. 

The fuel and air supply to the combustor are controlled using Alicat mass flow controllers (MFCs) which have an uncertainty $\pm$ (0.8$\%$ of reading + 0.2$\%$ of full scale). For all the experiments, we maintain the mass flow rate of fuel ($\dot m_{fuel}$) constant and continuously vary the mass flow rate of air ($\dot m_{air}$). We perform experiments at different $\dot m_{fuel}$ and different rates of change of $\dot m_{air}$. We control $\dot m_{air}$ by utilizing an NI DAQ. The lowest and highest values of Reynolds number in our experiments are 1.28 $\times10^4\pm$ 345 and 2.47$\times10^4\pm$ 594 respectively. 

By varying $\dot m_{air}$, we vary the equivalence ratio from 0.99 to 0.56 and simultaneously acquire experimental data. For all the experiments, we record temporal signals of acoustic pressure and global heat release rate at a sampling rate of 10 kHz by using an A/D card (NI-6143, 16 bit). The acoustic pressure signals were acquired using a piezoelectric pressure transducer (PCB103B02, uncertainty $\pm$ 0.15 Pa) that is mounted at a distance of 25 mm from the dump plane of the combustor. The pressure transducer is mounted on a T-joint and placed within a Teflon adapter. This mounting arrangement does not cause large changes in the amplitude of pressure oscillations nor the phase difference to significantly affect our analysis. We measure the global heat release rate fluctuations in the flame with a photomultiplier tube (PMT, Hamamatsu H10722-01) equipped with a CH$^*$ filter (wavelength of 430 nm and 12 nm FWHM). For selected experiments, we acquire spatially resolved high-speed CH* chemiluminescence images of the flame using a CMOS camera (Phantom - v12.1) at a sampling rate of 500 fps and a spatial resolution of 800 $\times$ 600 pixels. The camera was equipped with a ZEISS 50 mm camera lens at $f$/2 aperture outfitted with a CH$^*$ filter (wavelength of 430 nm and 12 nm FWHM).

We employ different designs of the bluff-body to create additional passages for the fluid to flow across the bluff-body. In the first design, we create four slots near the region attached to the shaft. We refer to this configuration as the inner slot (IS) bluff-body (Fig. \ref{Expt rig}c). For the second design, we create circular holes near the region that is attached to the shaft (Fig. \ref{Expt rig}d). The total area of these four circular holes is 113 mm$^2$, which is a quarter of the total area of the slots on the IS bluff-body. We refer to this configuration as the four hole (4H) bluff-body. The last design contains twenty three circular holes of diameter 5 mm, distributed over the bluff-body as shown in Fig. \ref{Expt rig}e. The total area of these circular holes is same as the total area of the slots in the IS bluff-body. We call this configuration as multiple-hole (MH) bluff-body.

\section{Results and Discussion}
\label{3. Results and discussions}

\subsection{Characterization of the phase transition at different operating conditions}
\label{Results:1}

First, we analyze the results obtained for experiments with a combustor length of 1100 mm, operating at $\dot m_{fuel}$ of 34 SLPM with the bluff-body positioned 45 mm from the dump plane and where $\dot m_{air}$ changes at a rate of 3 SLPM/s. We name this set of operating conditions and experimental configurations used in \citep{manikandan2021seeds} as the reference experimental configuration (REC) for the purpose of comparing the results. For all the subsequent figures, instead of discussing the plots with respect to $\dot m_{air}$, we describe the figures using the global equivalence ratio $\phi$. Even though the flame is only partially premixed, we use $\phi$ so that we can compare the results at different thermal power ratings. 

\begin{figure}[t!]
\centering
\includegraphics[width=1.0\textwidth]{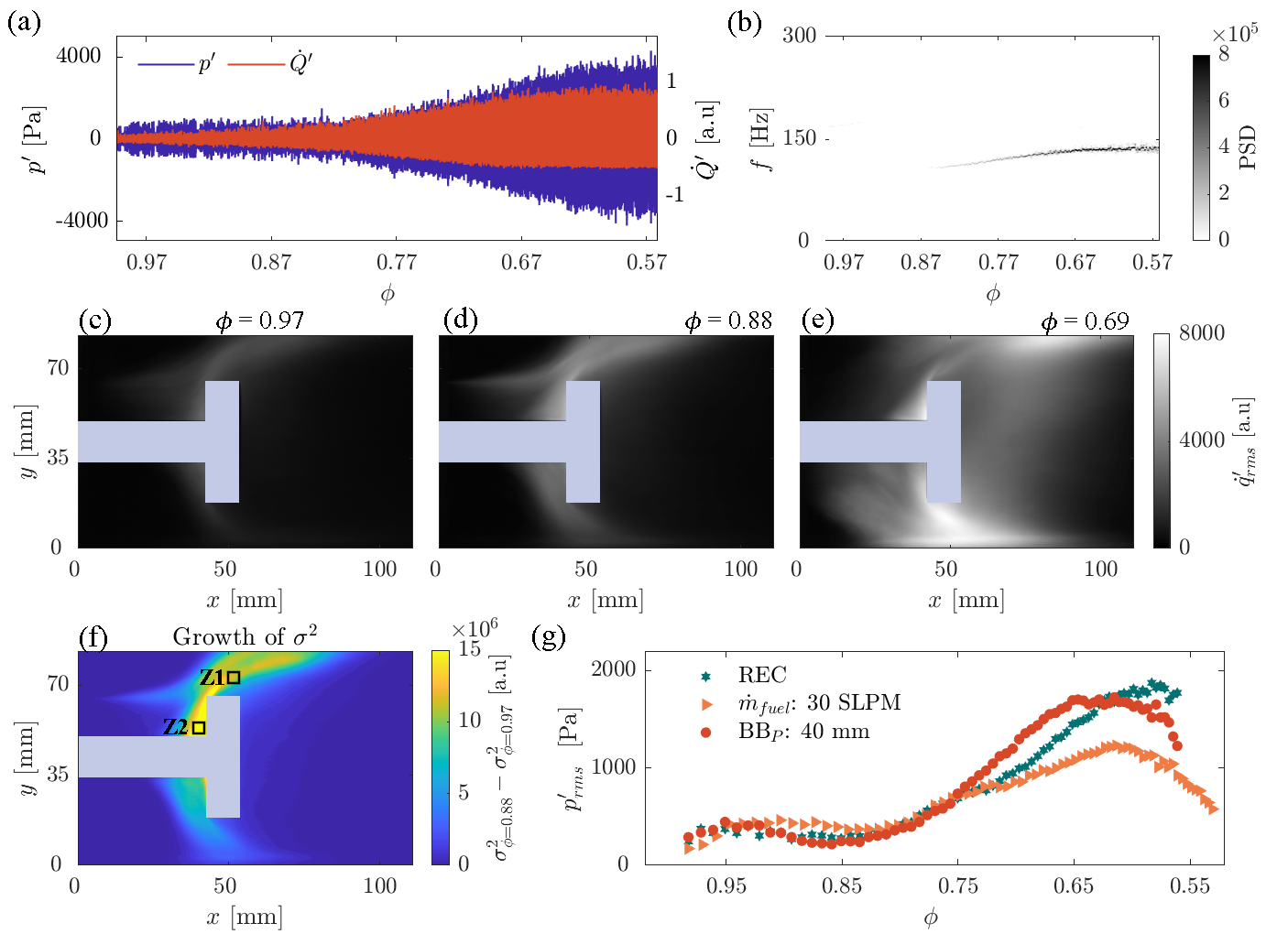}
\caption{Spatiotemporal dynamics with respect to the transition from combustion noise to thermoacoustic instability for the BL bluff-body at REC. (a) Time series of pressure oscillations $p^{\prime}$ and global heat release rate fluctuations $\dot Q^{\prime}$ with respect to $\phi$. (b) Evolution of power spectral density obtained using the STFT of $p^{\prime}$. $\dot q^{\prime}_{rms}$ at (c) $\phi$ = 0.97, (d) $\phi$ = 0.88 and (e) $\phi$ = 0.69. (f) Growth of $\sigma^2$ calculated as $\sigma^2_{\phi=0.88}$ - $\sigma^2_{\phi=0.97}$. (g) Variation in $p^{\prime}_{rms}$ for different operating conditions and combustor configurations and its described  in the legend. Reference experimental configuration (REC) refer to experiments with combustor length 1100 mm and position of the bluff-body (BB$_p$) 45 mm from the dump plane, operating at $\dot m_{fuel}$ 34 SLPM and rate of change of $\dot m_{air}$ at 3 SLPM/s. Other keys in the legend refers to the operating condition that is different from REC.}
\label{Pf-Qf-Freq}
\end{figure}

Decreasing $\phi$ results in a continuous growth in the amplitude of acoustic pressure ($p^{\prime}$) and global heat release rate oscillations ($\dot Q^{\prime}$), as shown in Fig. \ref{Pf-Qf-Freq}a. The short-time Fourier transform (STFT)  plot in figure \ref{Pf-Qf-Freq}b shows a dominant mode emerging near 100 Hz at $\phi$ $<$ 0.87 for $p^{\prime}$. As $\phi$ is reduced, the frequency of the dominant mode increases and reaches an asymptotic state for $\phi$ $<$ 0.67. At high values of $\phi$ (0.99 $\geq \phi \geq$ 0.88), $p^{\prime}$ and $\dot Q^{\prime}$ have dissimilar dominant modes with shallow bands (Pawar et al. \cite{pawar2017thermoacoustic}). 

Next, we analyze the local growth of heat release rate fluctuations $\dot q^{\prime}(x, y)$ as $\phi$ is reduced. The spatial distribution of the root mean square (mean is calculated for 1 second, which means 500 data points) of local heat release rate fluctuations $\dot q^{\prime}_{rms}$ is shown at three values of equivalence ratio, $\phi$ = 0.97 (Fig. \ref{Pf-Qf-Freq}c), $\phi$ = 0.88 (Fig. \ref{Pf-Qf-Freq}d) and $\phi$ = 0.69 (Fig. \ref{Pf-Qf-Freq}e). $\dot q^{\prime}_{rms}$ shows high strength around the bluff body and near the combustor walls. 

In the framework of phase transition theory, it is possible to analyze critical phenomena as the transition is approached by estimating the growth in variance of fluctuations ($\sigma^2$) of a system variable termed as pre-bifurcation noise amplification \citep{surovyatkina2005fluctuation}. We estimate the growth of $\sigma^2$ of the local heat release rate fluctuations from the difference between $\sigma^2$ at $\phi$ = 0.97 and $\phi$ = 0.88 (Fig. \ref{Pf-Qf-Freq}f). We use these two values of $\phi$ (also for subsequent sections) to calculate the growth of $\sigma^2$ prior to the occurrence of large-amplitude pressure oscillations (refer Fig. \ref{Pf-Qf-Freq}a) for the REC configuration. 

We mark two regions that have high growth of $\sigma^2$ using black squares in Fig. \ref{Pf-Qf-Freq}f. Z1 is a location of high heat release rate during the occurrence of thermoacoustic instability because of the impingement of large-scale coherent flow structures on the combustor wall. Z2 represents a stagnation point that stabilizes the flame because of the low velocity region at the corner between the shaft and the bluff-body. Both Z1 and Z2 exhibits critical phenomena prior to the transition to thermoacoustic instability, illustrated by the high growth in $\sigma^2$ (Fig. \ref{Pf-Qf-Freq}f). In our previous study, we referred to these zones as ``seeds" of the phase transition due to critical phenomena occurring at these locations, well before the appearance of large-amplitude pressure oscillations. After the critical transition in the local heat release rate fluctuations occurs at these zones, the strength of pressure oscillations ($p_{rms}$) increases rapidly as $\phi$ is reduced for REC as shown in Fig. \ref{Pf-Qf-Freq}g. 

We now characterize the transition using the temporal measurements of $p^{\prime}$ for changes in the bluff-body position, and $\dot m_{fuel}$. As the bluff-body position from the dump plane is reduced to 40 mm from 45 mm (REC), it appears that large-amplitude oscillations occur earlier compared to REC (See Fig. \ref{Pf-Qf-Freq}g). But at $\phi$ $<$ 0.6, the pressure oscillations with 45 mm bluff-body position (REC) is stronger than those with 40 mm. 

We also investigate changes in the transition with respect to the thermal power rating: $\dot m_{fuel}$. A lower $\dot m_{fuel}$ does not affect the $p^{\prime}_{rms}$ values with respect to REC till $\phi$ = 0.7 (Fig. \ref{Pf-Qf-Freq}g). For $\phi <$ 0.7, as expected $p^{\prime}_{rms}$ is lower in comparison to REC.

\subsection{Reducing growth of fluctuations at Z2 and suppressing the transition to thermoacoustic instability}
\label{Results:2}

\begin{figure}[t!]
\centering
\includegraphics[width=1.0\textwidth]{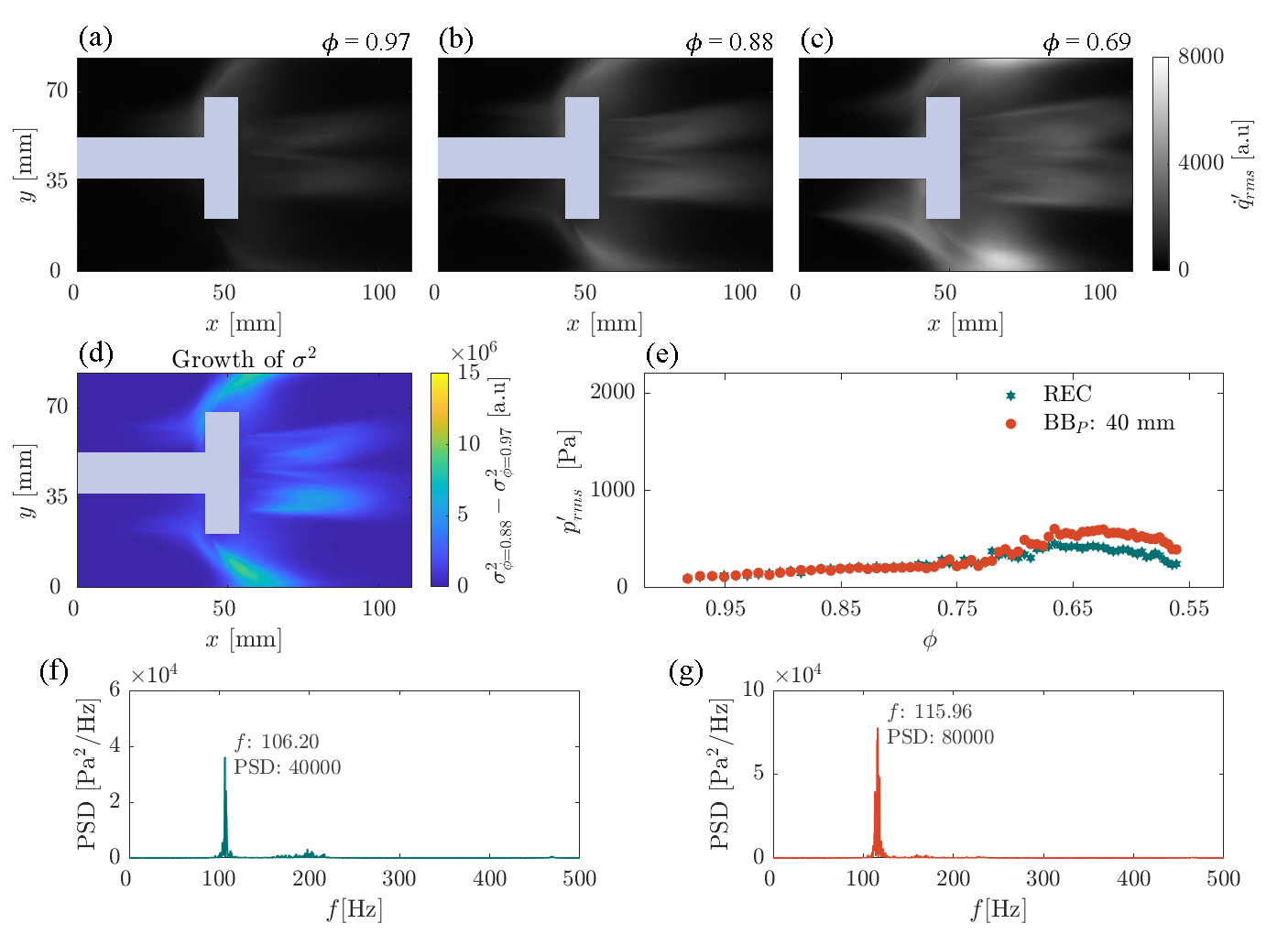}
\caption{Spatiotemporal dynamics for experiments conducted utilizing the IS bluff-body. $\dot q^{\prime}_{rms}$ at (a) $\phi$ = 0.97, (b) $\phi$ = 0.88 and (c) $\phi$ = 0.69 for REC. (d) Growth of $\sigma^2$: $\sigma^2_{\phi = 0.88}$ - $\sigma^2_{\phi = 0.97}$ for REC. (e) $p^{\prime}_{rms}$ for different operating conditions and combustor configurations with respect to $\phi$. Power spectral density of $p^{\prime}$ for (f) REC, (g) BB$_p$: 40 mm.}
\label{qrms_VOF-prms_Qrms_Freq_IS}
\end{figure}

The high growth in $\sigma^2$ at Z1 and Z2 (Fig. \ref{Pf-Qf-Freq}f) occurs for all the operating conditions that we have described above. In the current parametric study on the suppression of the phase transition to thermoacoustic instability, we will target Z2 and modify the flame distribution around the bluff-body. Utilizing the IS bluff-body design, we introduce passages on the bluff-body in the form of slots. These slots are introduced between a radius of 8 mm and 15 mm. Since the shaft radius is also 8 mm, we eliminate the stagnation zone at the corner between the bluff-body and the shaft in order to reduce the growth of local heat release rate fluctuations at Z2. We first analyze results with the IS bluff-body at REC. 

Figures \ref{qrms_VOF-prms_Qrms_Freq_IS}a-c shows the variation in the spatial distribution of $\dot q^{\prime}_{rms}$ as the equivalence ratio is reduced. The IS configuration results in higher strength of the local heat release rate oscillations downstream of the bluff-body (see Figs. \ref{qrms_VOF-prms_Qrms_Freq_IS}b, c) in comparison to the BL configuration. In comparison to the BL bluff-body at REC, we observe that the strength of the heat release rate oscillations ($\dot q^{\prime}_{rms}$) have reduced at the stagnation point (refer Figs. \ref{Pf-Qf-Freq}d, e and \ref{qrms_VOF-prms_Qrms_Freq_IS}b, c). The highest strength of $\dot q^{\prime}$ oscillations are observed near the combustor walls (see Fig. \ref{qrms_VOF-prms_Qrms_Freq_IS}c). 

Thus, the inner slot (IS) bluff-body design redistributes the heat release rate fluctuations around the bluff-body. This redistribution results in a reduction of growth of $\sigma^2$ at Z2 (Fig. \ref{qrms_VOF-prms_Qrms_Freq_IS}d). This low growth corresponds with a reduction of $\dot q^{\prime}_{rms}$ at Z2. On the other hand, near Z1, the growth of $\sigma^2$ is still high. Further, the wake region of the IS bluff-body also indicates a growth of $\sigma^2$. But importantly, the growth of $\sigma^2$ is not as high as the growth at Z2. It is crucial that other regions do not show high growth of $\sigma^2$ due to the change in the bluff-body design. Other regions showing high growth of $\sigma^2$ may mean the existence of other pathways for the transition to thermoacoustic instability.

The plot of $p^{\prime}_{rms}$ at REC shows the suppression of the global transition to thermoacoustic instability (Fig. \ref{qrms_VOF-prms_Qrms_Freq_IS}e). $p^{\prime}_{rms}$ remains low as $\phi$ is reduced to low values. In fact, at low values of $\phi$, there is 88$\%$ decrease in the $rms$ value of pressure oscillations in comparison to the $rms$ value obtained with BL bluff-body at REC. However, $p^{\prime}_{rms}$ at low values of $\phi$ with IS bluff-body is 0.45 kPa that is approximately 21$\%$ higher than the strength of pressure oscillations ($p^{\prime}_{rms}$ = 0.37 kPa) at high values of $\phi$ (combustion noise) with the BL bluff-body. The power spectral density of $p^{\prime}$ for REC at $\phi$ = 0.69 shows a shallow band centered around 106.2 Hz (Fig. \ref{qrms_VOF-prms_Qrms_Freq_IS}f). This dominant mode obtained using IS bluff-body at REC is at a considerably lower frequency than the dominant mode observed for the BL bluff-body at REC.

Irrespective of the changes in the thermal power loading ($\dot m_{fuel}$), there is a suppression of thermoacoustic instability when the IS bluff-body is used (shown in Supplementary Fig. S1). However, a higher reduction of the strength of pressure oscillations is observed for higher $\dot m_{fuel}$. Further, as the bluff-body is brought closer to the dump plane from 45 mm to 40 mm, the reduction in the strength of oscillations is not as large as that observed with 45 mm bluff-body position. For this bluff-body position of 40 mm, the PSD of $p^{\prime}$ at $\phi$ = 0.69 displayed in Fig. \ref{qrms_VOF-prms_Qrms_Freq_IS}h shows a shallow band centered around 116 Hz, which is higher when compared to the band corresponding to REC with IS bluff-body. However, this frequency is still lower than the narrow band observed for REC with BL bluff-body (Refer Fig. \ref{Pf-Qf-Freq}b).

\subsection{Effect of reducing the area of the slots on the bluff-body}
\label{Results:3}

\begin{figure}[t!]
\centering
\includegraphics[width=1.0\textwidth]{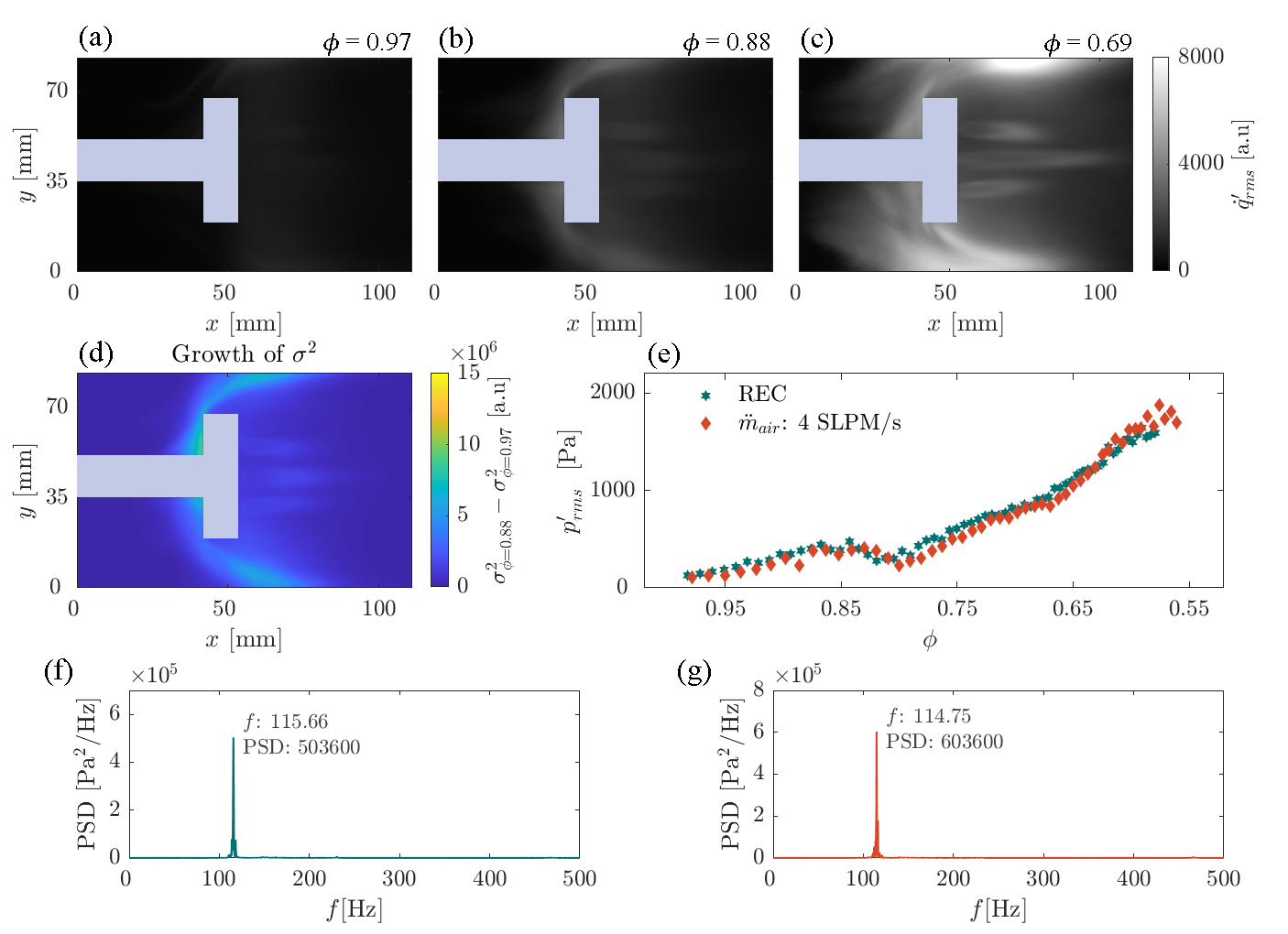}
\caption{Spatiotemporal dynamics for experiments conducted using the 4H bluff-body at REC. $\dot q^{\prime}_{rms}$ at (a) $\phi$ = 0.97, (b) $\phi$ = 0.88 and (c) $\phi$ = 0.69. Growth of $\sigma^2$: $\sigma^2_{\phi_2}$ - $\sigma^2_{\phi_1}$. $p^{\prime}_{rms}$ for different operating conditions and combustor configurations with respect to $\phi$. Power spectral density of $p^{\prime}$ for (f) REC, (g) rate of change of mass flow rate of air: $\ddot m_{air}$ = 4 SLPM/s.}
\label{qrms_VOF-prms_Qrms_Freq_FH}
\end{figure}

We investigate the effect of changes in the area of the passage by the four-hole (4H) bluff-body. Analysis of the local heat release rate fluctuations prove that for the 4H bluff-body, at Z1 and Z2, i) higher $\dot q^{\prime}_{rms}$ exists compared to IS bluff-body (Figs. \ref{qrms_VOF-prms_Qrms_Freq_FH}a-c) and ii) high growth of $\sigma^2$ occurs (Fig. \ref{qrms_VOF-prms_Qrms_Freq_FH}d). In addition, we observe a small growth of $\sigma^2$ in the wake of the 4H bluff-body due to secondary flames, as observed with the IS bluff-body. 

When comparing the experimental results at REC between 4H and BL bluff bodies, we do not observe any large changes in the transition from low-amplitude pressure fluctuations to high-amplitude pressure oscillations (Fig. \ref{qrms_VOF-prms_Qrms_Freq_FH}e). A gradual growth in $p^{\prime}_{rms}$ occurs when using REC (Fig. \ref{qrms_VOF-prms_Qrms_Freq_FH}e). The strength of $p^{\prime}$ at low $\phi$ is also similar in magnitude with the strength obtained for BL bluff-body. However, the power spectral density of $p^{\prime}$ at $\phi$ = 0.69 shows a sharp peak centered around 115.7 Hz (Fig. \ref{qrms_VOF-prms_Qrms_Freq_FH}f), which is significantly lower than the the dominant mode observed with the BL bluff-body (Fig. \ref{Pf-Qf-Freq}b) for the same combustor length. Thus, from the results obtained with both IS and 4H bluff bodies, it is evident that the introduction of slots has shifted the dominant frequency to a lower value for the experiments performed at REC. 

When operating at a higher rate of change of $\dot m_{air}$, the change in the transition is negligible in comparison to the $p^{\prime}_{rms}$ for REC (Fig. \ref{qrms_VOF-prms_Qrms_Freq_FH}e). The dominant frequency at 0.69 is at 114.7 Hz (Fig. \ref{qrms_VOF-prms_Qrms_Freq_FH}g). Thus, multiple experiments with the 4H bluff-body show that the onset of thermoacoustic instability is not suppressed. The reduction in the amplitude of pressure oscillations for the 4H configuration is negligible with respect to the BL bluff-body.

\subsection{Effect of change in the distribution of the bluff-body passage area}
\label{Results:4}

\begin{figure}[t!]
\centering
\includegraphics[width=1.0\textwidth]{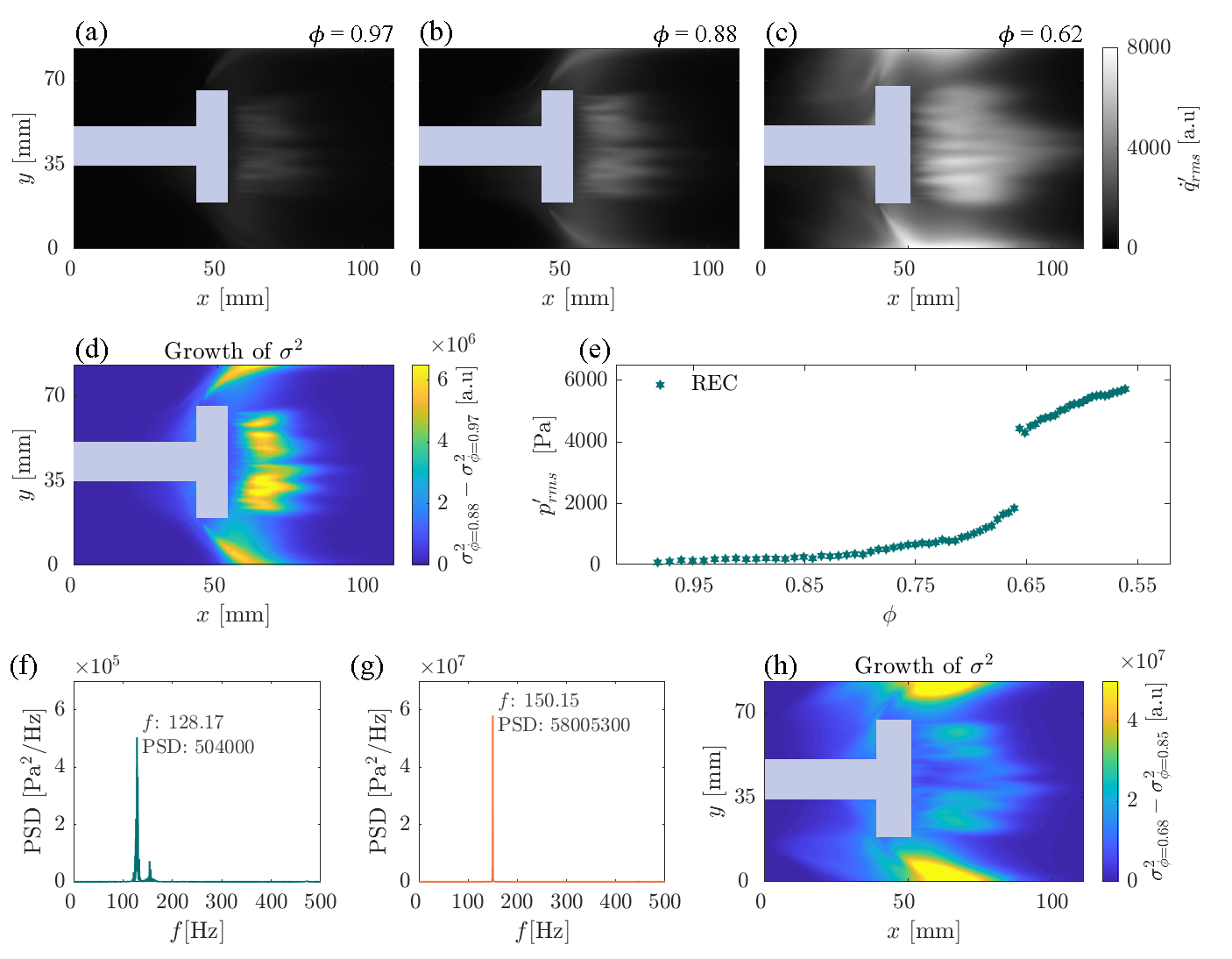}
\caption{Spatiotemporal dynamics for experiments conducted utilizing the MH bluff-body at REC. $\dot q^{\prime}_{rms}$ at (a) $\phi$ = 0.97, (b) $\phi$ = 0.88 and (c) $\phi$ = 0.62. (d) Growth of $\sigma^2$ for the primary bifurcation: $\sigma^2_{\phi = 0.88}$ - $\sigma^2_{\phi = 0.97}$. (e) $p^{\prime}_{rms}$ for different operating conditions and combustor configurations with respect to $\phi$. Power spectral density of $p^{\prime}$ for (f) primary bifurcation at $\phi$ = 0.69 and (g) secondary bifurcation at $\phi$ = 0.62. (h) Growth of $\sigma^2$ for secondary bifurcation: $\sigma^2_{\phi = 0.68}$ - $\sigma^2_{\phi = 0.75}$.}
\label{qrms_VOF-prms_Qrms_Freq_MultiHole}
\end{figure}

We now investigate the effect of change in the distribution of the passages on the bluff-body. To that end, we utilize the MH bluff-body. In Figs. \ref{qrms_VOF-prms_Qrms_Freq_MultiHole}a-c, we observe high strength of $\dot q^{\prime}_{rms}$ at Z1 and downstream of the MH bluff body for REC. As seen in Fig. \ref{qrms_VOF-prms_Qrms_Freq_MultiHole}d, with the MH bluff-body, the growth of $\sigma^2$ still exists at Z2, although it is low. Higher growth of $\sigma^2$ downstream of the bluff-body can be attributed to the large heat release rate fluctuations of the multiple secondary flames anchored on the bluff-body surface.

Using the MH bluff-body, we observe a similar transition from combustion noise to thermoacoustic instability to that of BL bluff body till $\phi$ = 0.75 (refer Figs. \ref{qrms_VOF-prms_Qrms_Freq_MultiHole}e and \ref{Pf-Qf-Freq}g). The $p^{\prime}_{rms}$ for REC (679 Pa) is slightly higher compared to the $p^{\prime}_{rms}$ obtained with the BL bluff-body at REC (644 Pa). The power spectral density of $p^{\prime}$ at $\phi$ = 0.69 show the dominant frequency at 128 Hz (Fig. \ref{qrms_VOF-prms_Qrms_Freq_MultiHole}f). This dominant frequency is still lower than the dominant frequency obtained with the BL bluff-body (Fig. \ref{Pf-Qf-Freq}b) for the same combustor length, but closest to it as compared to the values obtained with IS and 4H bluff bodies. Thus, suppression of the transition does not occur for the MH bluff-body. 

Interestingly, as the equivalence ratio is reduced further to less than 0.66, an additional very sharp transition occurs as seen in Fig. \ref{qrms_VOF-prms_Qrms_Freq_MultiHole}e. The increase in the strength of pressure oscillations is rather catastrophic, increasing to almost 300 $\%$ of the $p^{\prime}_{rms}$ value after the primary bifurcation. This secondary bifurcation is observed only for the MH bluff body. The dominant frequency increases to 150 Hz for the secondary bifurcation at $\phi$ = 0.62 (Fig. \ref{qrms_VOF-prms_Qrms_Freq_MultiHole}g). 

This secondary bifurcation highlights the challenge in suppressing the onset of thermoacoustic instability without introducing other routes for the dangerous transition to occur. Although the reason behind introducing the slots on the bluff-body is to reduce the growth of local heat release rate fluctuations upstream of the bluff-body at the stagnation zone, not all designs result in a reduction of fluctuations and hence the suppression of thermoacoustic instability. In the case of the MH bluff-body, the highest growth of local heat release rate fluctuations is not observed at the stagnation point, but rather downstream of the bluff-body as a result of the secondary flames (Fig. \ref{qrms_VOF-prms_Qrms_Freq_MultiHole}h). The appearance of new regions with high growth of heat release rate fluctuations indicates new pathways through which thermoacoustic instability emerges. Thus, the MH configuration does not suppress thermoacoustic instability but in fact enhances the strength of pressure oscillations at certain conditions.

\section{Conclusion}
\label{Conclusions}

In this study, we analyzed the results of a suppression strategy to reduce the growth of local heat release rate fluctuations to prevent the transition from combustion noise to thermoacoustic instability. Our experimental investigation suggest that certain configurations of perforations on the bluff-body ensures reduction in the growth of local heat release rate fluctuations upstream of the bluff-body and hence aids in suppressing the onset of thermoacoustic instability. We observed that, in general, suppression of the onset of thermoacoustic instability is possible for such a passive control strategy. The success of this strategy shows that passive control can be designed based on the growth of local heat release rate fluctuations well before the occurrence of thermoacoustic instability.

However, the reduction in the amplitude of pressure oscillations is lower if the bluff-body is brought closer to the dump plane. The less effective suppression for this configuration using the IS bluff body shows the need to analyze the growth of local $\sigma^2$ when the bluff-body position is changed. Further, we need to have an optimization strategy to find the appropriate area of passage with respect to the position of the bluff-body such that the growth of local heat release rate fluctuations can be reduced at the stagnation zone.

Most importantly, a mere introduction of slots on the bluff-body does not result in the suppression of the transition to thermoacoustic instability. We observed that the suppression of the transition does not occur if the area of the slots is changed to a lower value (4H bluff-body) compared to the IS bluff body. 4H bluff-body does not cause significant reduction in the local heat release rate fluctuations at the stagnation zone and instead causes high amplitude pressure oscillations. We also found that an inappropriate distribution of the slots could result in a secondary bifurcation, leading to a catastrophic increase (300 $\%$) in the amplitudes of pressure oscillations (MH bluff-body). We found that the secondary bifurcation occurs via the growth of local heat release rate fluctuations at locations downstream of the bluff-body. These results highlight the challenge of suppressing the onset without initiating new pathways to transition to thermoacoustic instability. Thus, our study introduces an optimization strategy for passive control of thermoacoustic instability using spatiotemporal data obtained far away from high-amplitude thermoacoustic oscillations.

\section*{Acknowledgments}
This research is funded by the IoE initiative (SB/2021/0845/AE/MHRD/002696), IIT Madras, and Department of
Science and Technology (Grant no: CRG/2020/003/051), Government of India. N.B.G acknowledges the financial support of the East Africa Peru India Climate Capacities project (18$\_$II$\_$149$\_$Global$\_$A$\_$Risikovorhersage) funded by the Federal Ministry for the Environment, Nature Conservation and Nuclear Safety and the International Climate Initiative. E.S acknowledges the financial support by RFBR, project number 20-07-01071. The authors would like to express their gratitude to Mr. Midhun P.R., Mr. Anand S., and Mr. Thilagaraj S. of Aerospace department, IIT Madras for their assistance in conducting the experiments.

\section*{Supplementary material}
Supplementary material is submitted separately.

\bibliography{manuscript.bib}

\end{document}